\documentclass[prl,twocolumn,aps,showpacs,superscriptaddress,]{revtex4-2} 

\usepackage{amsfonts}
\usepackage{amsmath}
\usepackage{graphicx}
\usepackage{dcolumn}
\usepackage{bm}
\usepackage{xcolor} 
\usepackage{mathtools}

\usepackage{academicons}
\usepackage{orcidlink}
\definecolor{orcidlogocol}{HTML}{A6CE39}

\begin{document}

\title{\LARGE \textbf{Metabolic scaling, von Bertalanffy growth\\ and an exponent equation}}

\author{\orcidlink{0000-0002-0062-3377}Hana Krakovsk\'{a} }
\email[Corresponding author:]{hana.krakovska@savba.sk}
\affiliation{Institute of Measurement Science, Slovak Academy of Sciences, D\'{u}bravsk\'{a} cesta 9, Bratislava, Slovakia}
\affiliation{Institute of the Science of Complex Systems, Center for Medical Data Science, Medical University of Vienna, Spitalgasse 23, Vienna, 1090, Austria}
\affiliation{Complexity Science Hub, Metternichgasse 8, Vienna,
1030, Austria}

\author{\orcidlink{0000-0002-6887-7736}Klaus Stiefel}
\affiliation{Neurolinx Research Institute, La Jolla, CA, USA}
\affiliation{Silliman University Angelo King Center for Research \& Environmental Management, Dumaguete City, Philippines}
\author{\orcidlink{0000-0002-4045-9532}Rudolf Hanel} 
\affiliation{Institute of the Science of Complex Systems, Center for Medical Data Science, Medical University of Vienna, Spitalgasse 23, Vienna, 1090, Austria}
\affiliation{Complexity Science Hub, Metternichgasse 8, Vienna,
1030, Austria}

\begin{abstract}
In this work, we 
interpret developmental growth
as a metabolic energy allocation problem and link the von Bertalanffy growth model to metabolic energy investments into the growth channel.
Using a framework that specifies how metabolic energy is allocated among baseline maintenance, growth, and other processes, we analyse the resulting growth allocation patterns and derive direct relationships between key scaling exponents: the mass-growth exponent $\alpha_{\rm gr}$, the length-based exponent $\alpha$, the metabolic scaling exponent $\alpha_{\rm mbl}$, and the geometric exponent $\beta$, which describes the mass-length relationship.
These exponents
determine the metabolic investment exponent
$\alpha_{\rm met}=\alpha_{\rm gr}-\alpha_{\rm mbl}$, which controls the
qualitative behaviour of the growth-allocation function $p_{\rm gr}$.

Requiring the inferred allocation fraction to remain biologically feasible,
$0\leq p_{\rm gr}\leq1$, we derive 
constraints on developmental velocity
and characteristic mass scales.  This provides a physical, energy-based
interpretation of phenomenological growth curves and clarifies how metabolic
scaling, geometric scaling, and growth dynamics are interrelated within a single
allocation framework.  
\end{abstract}
 \maketitle
\section{Introduction}
The growth of an animal from the fertilised oocyte to adult size is a fundamental biological process. Growth rate and typical adult size are central to understanding both ontogenetic development and ecological roles of individual organisms.
Quantitative models of animal growth are therefore indispensable for developing
a coherent understanding of growth phenomenology.
A physical, energy-based framework that accounts for how organisms invest energy into essential processes allows us to extend the interpretative range of such models.
Our aim here is to connect expected animal growth trajectories
with the energy intake and expenditure of organisms in different life phases
and to characterise the relationships between growth, geometrical species features and metabolic exponents.

Historically, various growth models have been introduced~\cite{kearney2021}. One of the most well-known is the von Bertalanffy growth model~\cite{bertalanffy1957} first introduced by P\"{u}tter~\cite{putter1920}, where the rate of mass gain (anabolism) is proportional to the surface area ($\alpha_{\rm gr}$ is usually taken to be $2/3$), and the rate of catabolism (mass loss) is proportional to total mass ($\eta$ is usually taken to be $1$):
\begin{equation}\label{eq:vb_de}
    \dot m=p^*\ m^{\alpha_{\rm gr}}-q^*\ m^\eta \ ,
\end{equation}
where $m$ is the mass of the organism, $\dot m=dm/dt$ is the derivative of $m$ with respect to age $t.$  Constants $p^*,\alpha_{\rm gr},q^*,\eta$ are non-negative, where $p^*$ is the anabolism constant, $\alpha_{\rm gr}$ the scaling exponent of anabolism, $q^*$ the catabolism constant and $\eta$ the exponent for mass scaling of catabolism.

The von Bertalanffy model is widely favoured for its simplicity, owing to its small number of fitting parameters and the availability of an explicit solution derived from the differential equation~\eqref{eq:vb_de}.
Additionally, it generally provides good fits to empirical data, with fit parameters for multiple species recorded in online databases such as FishBase~\cite{fishbase}. Growth rates differ significantly between species, and this variation is best explained by the order-of-magnitude differences in body size~\cite{stiefel2022}.

The growth curves are often typical for a species, while individual growth patterns 
may deviate significantly from these averages. For instance, even the monotonicity of an individual's growth curve in general is not 
guaranteed (e.g.~organisms may build up and later use up fat reserves). Moreover, the very early growth phase (the so-called \textit{lag phase}) is often accompanied by a loss of mass at some point rather than an increase (e.g.~the 
egg is heavier than the hatchling~\cite{halbersleben1922}). Similarly, a decrease in body mass towards the end of the natural life span is also not atypical.

On top of growth, an individual 
has to allocate energy to various other essential tasks 
including
baseline metabolic energy requirements, immune system related processes, reproduction, 
energy reserve management and others. 
Environmental conditions further bias such energy investments.
Under harsh conditions, differences in energy allocation and limited energy availability can result in reduced growth~\cite{dmitriew2011}.

Furthermore, even if we accept the von Bertalanffy growth model as an effective description of a species' \textit{growth program} rather than an individual's growth trajectory, a long-standing debate regarding the exact value of the exponent $\alpha_{\rm gr}$ in the model remains. This exponent is often assumed to be linked to the metabolic scaling exponent. However, the metabolic scaling exponent itself has been debated for over a century and measured across a wide range of taxa~\cite{kearney2021}. 

Early empirical work by Rubner supported a scaling exponent of $\sim 2/3,$ consistent with a theory based on the surface law~\cite{sarrus1838,rubner1883}. Later, $\sim 3/4$ metabolic scaling in mammals was reported in data-driven studies by Kleiber~\cite{kleiber1932} and Brody~\cite{brody1946}. Many studies have analysed data for both intraspecific and interspecific metabolic scaling (see a broad review by Glazier~\cite{glazier2005}). Some support the baseline metabolic rate exponent of $3/4$~\cite{savage2004,peters1986}, Dodds et al.~\cite{dodds2001}, for instance, are not able to reject the null hypothesis that the exponent is $2/3$ in favour of $3/4,$ other studies report the $2/3$ exponent (e.g.~\cite{heusner1982,white2003} in mammals). Feldman and McMahon~\cite{feldman1983} argue that the intra-specific exponent is around $2/3$ while inter-specific is around $3/4.$ Additionally, some authors claim there may not even be a single universal exponent or that a spectrum of exponents exists throughout the species' life~\cite{glazier2005,glazier2022}.

Von Bertalanffy~\cite{bertalanffy1957} specified three categories of metabolic scaling types which determine the growth exponent in the model. For fish, certain invertebrates, and partially mammals, he suggested a growth exponent of $\alpha_{\rm gr} \sim 2/3$, for insects, he proposed $\alpha_{\rm gr} = 1$, corresponding to exponential growth that stops at metamorphosis. The third group included species for which the exponent falls between these two extremes.

Alternatively, West et al.~\cite{west1997} proposed a model for metabolic scaling based on fractal resource distribution networks in organisms, predicting that metabolic rate $P$ scales with body mass $m$ as $P \propto m^{3/4}.$ In light of this, they proposed an ontogenetic growth model of the same form as von Bertalanffy's, but with a fixed growth exponent of $\alpha_{\rm gr} \sim \frac{3}{4}$~\cite{west2001,hou2008} (also see~\cite{makarieva2004}). More recently, a growth model with $\alpha_{\rm gr} = \frac{2}{3}$ and $\eta = \frac{3}{4}$ has been suggested~\cite{sibly2020}. 

Additionally, a broad body of literature directly fits the von Bertalanffy growth model to empirical growth data (see e.g.~studies on fish species~\cite{lee2020}, birds~\cite{tjorve2010} and fish and non-fish species~\cite{renner2018}). While debate over the true values of the metabolic and growth scaling exponents continues, the von Bertalanffy model nonetheless provides satisfactory empirical fits to observed growth patterns.

Here, we analyse a model of growth energy allocation in which the metabolic scaling exponent, which describes total energy expenditure, need not coincide with the exponent governing species-level growth. 

Our approach differs from classical models by von Bertalanffy, West et al.~\cite{west2001}, and Dynamic Energy Budget (DEB) theory~\cite{kooijman2010}, in which growth is commonly formulated as a balance between anabolism and catabolism, anabolism and maintenance, and assimilation, maintenance, reserve dynamics and other processes, respectively.

Here, we treat growth as a species-specific energy allocation priority relative to other physiological processes, such as reproduction, immune function and others.
We ask what growth-allocation function is required for a given von Bertalanffy trajectory to be compatible with an independently specified metabolic scaling law. 

In this interpretation, the empirically fitted growth curve acts as a
constraint on energy allocation. The allocation function $p_{\rm gr}(m)$ then
describes the fraction of surplus metabolic rate that must be devoted to
growth in order to reproduce the observed growth trajectory. Requiring this
fraction to remain biologically feasible, $0\leq p_{\rm gr}\leq1$, yields
constraints on the relationships among metabolic scaling, growth scaling,
geometric scaling, and developmental velocity.

An additional layer of complexity is added by the fact that in many empirical contexts, animal growth is described using a length-based formulation of the von Bertalanffy model rather than a mass-based one. Therefore, we also compare the length-based and mass-based growth models via a mass-length geometrical relationship and establish explicit links between their respective scaling exponents.
\section{Methods}
We will consider an energy allocation problem. Let us assume that baseline metabolic requirements are given~by
\begin{equation}\varepsilon_0(m)=\kappa_0\left(\frac{m}{m_0}\right)^{\alpha_{\rm mbl}},
\end{equation}
for a suitable mass unit $m_0$ and the baseline power $\kappa_0$ at scale $m_0$.

 We define the baseline metabolic requirements as the energy per unit time required to maintain baseline physiological processes that sustain life 
 (such as base metabolism, biomolecule turnover, thermogeneration in the case of mammals).
We assume the total metabolic rate for the species is some factor of the baseline metabolic requirements, i.e.~$\varepsilon = (1+\delta)\varepsilon_0$, where $\delta > 0$.
Note that $\delta$ may depend on age or environmental conditions, for example, it may be lower under competitive stress, or vary with life stage, lifestyle, or locomotion requirements. For simplicity, however, we will treat $\delta$ as a constant in the present analysis.
We therefore adopt the same scaling exponent for the total metabolic rate $\varepsilon$ as for the baseline metabolic requirements. 

The surplus metabolic energy per unit time is given by $\Delta\varepsilon_f=\delta \varepsilon_0.$ 
It can be invested into various tasks, one of them being growth
\[\Delta\varepsilon_{\rm gr}=p_{\rm gr}\,\Delta\varepsilon_f \ ,\] with $0\leq  p_{\rm gr}\leq 1$ describing the fraction of surplus metabolic energy per unit time that an organism invests into growth with the remainder \(1 - p_{\rm gr}\) allocated to other processes, such as reproduction, immune system and others. Assuming the species grows according to the von Bertalanffy law, we can estimate what portion of energy the species spends on growth throughout its growing cycle. This partitioning then provides a basis for analysing species’ life-history strategies.

In the following, we estimate the growth investment ratio $p_{\rm gr}$ and investigate the relationships among the exponents governing mass-based growth, length-based growth, geometric scaling, and metabolic investment (see Table~\ref{tab:symbols}). Instead of modelling growth in
terms of mass, the von Bertalanffy model~\eqref{eq:vb_de} is commonly formulated in terms of length $L$
\begin{equation}\label{eq:dl}
    \dot L = p\, L^\alpha - q\, L \,,
\end{equation}
where $\dot L=dL/dt$ is the derivative of $L$ with respect to age $t$, $p$ and $q$ are species-specific constants of anabolism and catabolism, respectively, and $\alpha$ is a scaling parameter. 
The differential equation~\eqref{eq:dl} can be rewritten as 
   \begin{equation}\label{eq:dl2}
\dot{L}=K\ \mu_L \left( \frac{L}{L_0} \right)^{\alpha} 
\left( 1 - \left( \frac{L}{L_\infty} \right)^{1-\alpha} \right) \ ,
\end{equation}
where $L_0$ is a positive constant (not necessarily equal to the initial length $L(0)$),
    $\mu_L=L_\infty\frac{\left( L_0/L_\infty \right)^{\alpha}}{1-\alpha},$
${L_\infty=\left(\frac{p}{q}\right)^{\frac{1}{1-\alpha}},}$ and $K\equiv q(1-\alpha)$ is the developmental velocity.
The solution to Eq.~\eqref{eq:dl} where $\alpha<1$ is
\begin{equation}
    \frac{L}{L_{\infty}}=\left(1-\left(1-\left( \frac{L(0)}{L_\infty} \right)^{1-\alpha}\right)\exp(-Kt)\right)^{\frac{1}{1-\alpha}} \ .
    \label{eq:growthfunctions}
\end{equation}
In empirical studies on fish growth dynamics, $\alpha$ is commonly set to $0$, in which case we obtain the widely used von Bertalanffy growth function
\begin{equation}
    L(t)=L_{\infty}\left(1-e^{-K(t-t_0)}\right) \ ,
\end{equation}
where in terms of the parametrisation of Eq.~\eqref{eq:growthfunctions} ${t_0=\frac{\ln(1-L(0)/L_{\infty})}{K}.}$

Next, we look at the connection between mass-based-growth exponent $\alpha_{\rm gr}$ and length-based exponent $\alpha.$ The relationship between length $L$ and mass $m$ is approximately given by $m = \rho L A,$ where $\rho$ is the density (which is assumed to be constant) and $A \sim R^2$ is the average cross-sectional area of the individual (with $ R$ being the radius of the cross-section). Different species exhibit scaling relations of the form
\begin{equation}\label{eq:lengthvsm}
L = L_0 \left( \frac{m}{m_0} \right)^\beta,
\end{equation}
where $\beta$ depends on the body geometry of the species (this formulation is mainly used for fish, for which large datasets are available). Commonly, the scaling in literature is provided as $m=aL^b$ where $b=1 /\beta.$

Suppose that relationship~\eqref{eq:lengthvsm} is true, then the corresponding differential equation to Eq.~\eqref{eq:dl2} for mass $m$ is 
\begin{equation}\label{eq:dm_1}
\dot{m} =K\ \mu_m \left( \frac{m}{m_0} \right)^{\alpha_{\rm gr}}\left( 1 - \left( \frac{m}{m_\infty} \right)^{1-\alpha_{\rm gr}} \right) \ ,
\end{equation}

with $\mu_m=\frac{m_0}{L_0\beta}\mu_L$ and 
\begin{equation}\label{eq:grvsalpha_beta}
 \boxed{\alpha_{\rm gr}=1-\beta(1-\alpha)} \ .
\end{equation}
\begin{table}
\centering
\renewcommand{\arraystretch}{1.4}
\begin{tabular}{lllp{10.5cm}}
\hline
\textbf{Symbol} & \textbf{Name / Description}\\
\hline
$\varepsilon_0(m)$ & \rule{0pt}{2.5ex}Baseline metabolic requirements &\\
$\alpha_{\rm mbl}$ & Baseline metabolic scaling exponent & \\
$\alpha_{\rm gr}$ & Mass-based von Bertalanffy growth exponent  \\
$\alpha$ & Length-based von Bertalanffy growth exponent  \\
$\beta$ & Geometry exponent &  \\
$\alpha_{\rm met}$ & Metabolic investment exponent & \\
$p_{\rm gr}$ & Fraction of surplus metabolic rate for growth &   \\
$K$ & Developmental velocity  \\
\hline
\end{tabular}
\caption{Most important symbols and parameters used in the model.}
\label{tab:symbols}
\end{table}
Next, we relate metabolic rate to growth by assuming that the growth rate is proportional to the allocated metabolic energy per unit time, with proportionality constant $\gamma$ representing the conversion of metabolic energy into new biomass. This yields
\begin{equation}\label{eq:dm_2}
    \dot m = \gamma\, p_{\rm gr}\, \delta\, \varepsilon_0(m) \,.
\end{equation}
By comparing the von Bertalanffy equation~\eqref{eq:dm_1} and the metabolic equation~\eqref{eq:dm_2}, we can determine the metabolic energy per unit time investment ratio $p_{\rm gr}(m)$ as
\begin{equation}\label{eq:p_gr}
\boxed{
    p_{\rm gr}=\frac{K\mu_m}{\gamma\delta\kappa_0}\left( \frac{m}{m_0} \right)^{\alpha_{\rm gr}-\alpha_{\rm mbl}} 
\left( 1 - \left( \frac{m}{m_\infty} \right)^{1-\alpha_{\rm gr}} \right)}\ ,
\end{equation}
where we denote the metabolic investment exponent as $\alpha_{\rm met}\equiv\alpha_{\rm gr}-\alpha_{\rm mbl}.$

As noted above, to avoid introducing variability in the growth exponent, we assume $\delta$ is constant and age-independent.
\begin{figure*}
    \centering
        \includegraphics[height=5.4cm,trim={0.3cm 9.5cm 1.5cm 8.5cm},clip]{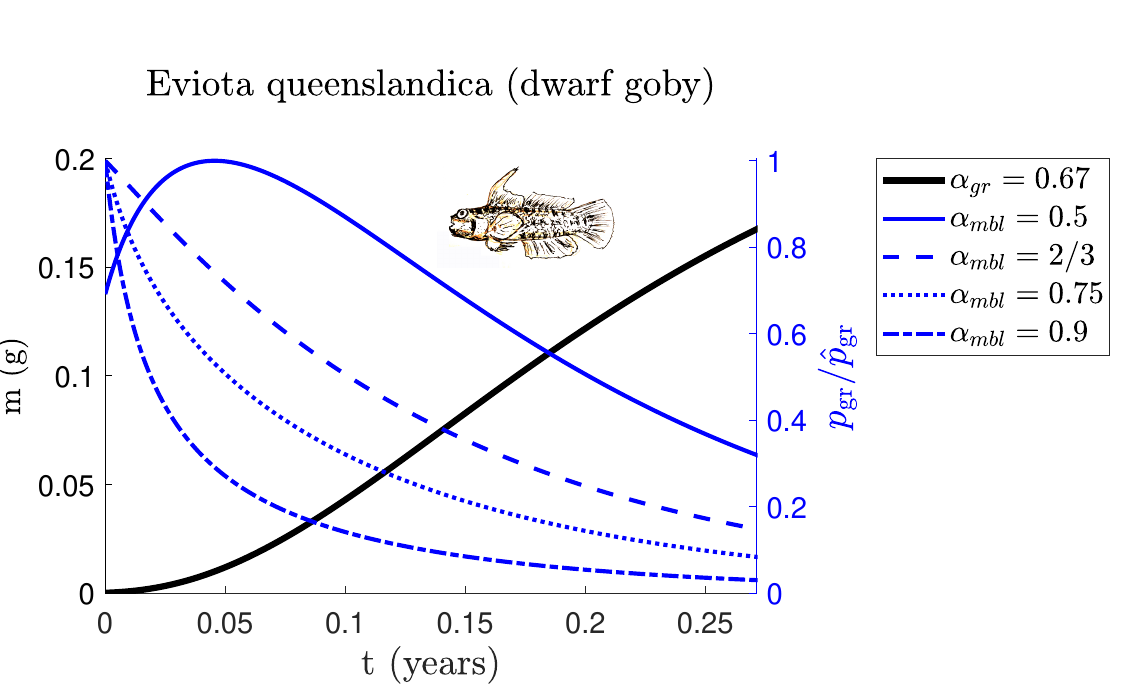}
           \includegraphics[height=5.4cm,trim={0.3cm 9.5cm 1.5cm 8.5cm},clip]{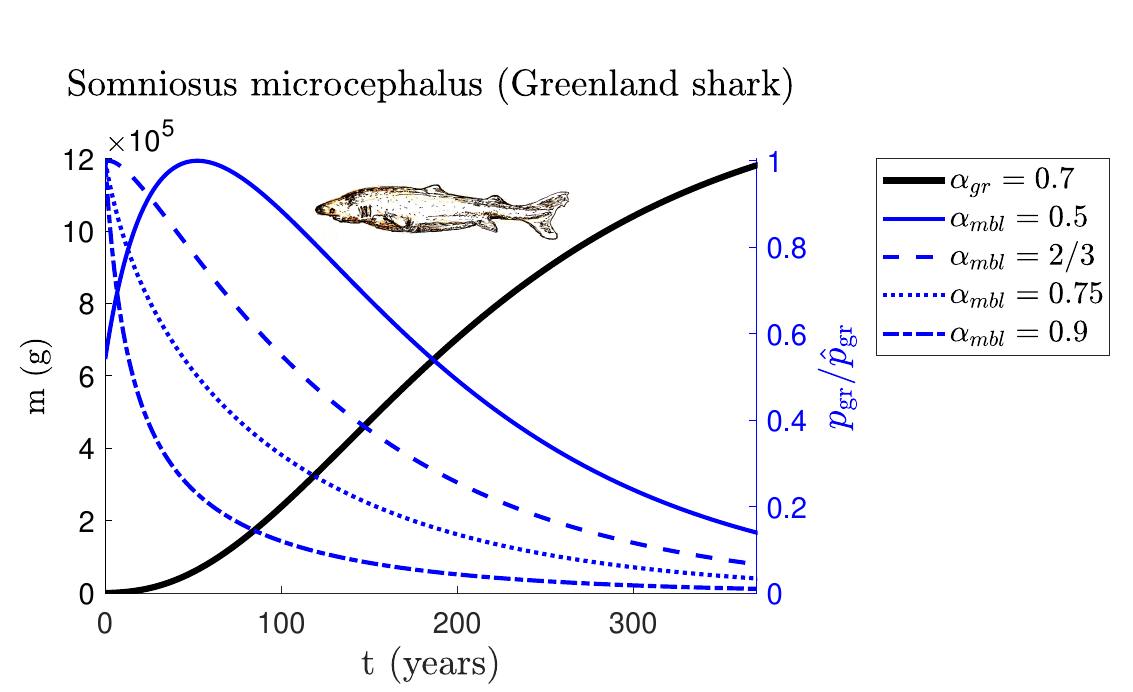}
    \caption{
    Growth curve (black) and scaled investment ratios $p_{\rm gr}/\hat{p}_{\rm gr}$ (blue), where $\hat{p}_{\rm gr}$ is the maximal relative growth investment, for different assumed values of metabolic scaling $\alpha_{\rm mbl}$ of fish species. On the left, indeterminate growth of \textit{Eviota queenslandica} ($\alpha_{\rm gr}=0.67$) and different $p_{\rm gr}$ growth allocation functions depending on the given metabolic scaling exponent $\alpha_{\rm mbl}$ (see the legend). On the right, the same for \textit{Somniosus microcephalus} ($\alpha_{\rm gr}=0.7$).
    The curves of $p_{\rm gr}$ corresponding to $\alpha_{\rm mbl}=0.5$ have a maximum, while the other curves are monotonically decreasing. Although the magnitude of $p_{\rm gr}$ varies with parameter choice in the prefactor $\gamma\delta\kappa_0$ (see Eq.~\eqref{eq:p_gr}), its qualitative shape, shown here mainly for illustrative purposes, reflects the relative growth allocation priorities over time. The fish illustrations were created by the authors.}
    \label{fig:pgr_fig}
\end{figure*}

Next, let us analyse the monotonicity of the function~$p_{\rm gr}.$ The interior extremum at the point $\hat{m}$ is found from the condition
$\frac{dp_{\rm gr}}{dm}(\hat{m}) = 0 ,$ which leads to
\begin{equation}
\label{eq:pgr_max}
\left(\frac{\hat m}{m_\infty}\right)^{1-\alpha_{\rm gr}}=\frac{\alpha_{\rm gr}-\alpha_{\rm mbl}}{1-\alpha_{\rm mbl}} \ ,
\end{equation}
which is feasible when $\alpha_{\rm gr} >\alpha_{\rm mbl}$ and $m(0)< \hat m$ (otherwise $p_{\rm gr}$ is a monotonically decreasing function of mass with a maximum at $m(0)$).
Because $0 \le p_{\rm gr}(m) \le 1$ must hold for all masses, the condition must be satisfied at the maximal value of $p_{\rm gr}$. If the interior maximum $\hat m$ from Eq.~\eqref{eq:pgr_max} lies within the admissible domain, enforcing $p_{\rm gr}(\hat m)\le 1$ leads to the following bound on the developmental velocity $K$:
\begin{equation}
    K\leq\gamma\,\delta\kappa_0 
    Q \ ,
\end{equation}
where
\begin{equation}
     Q=\left(\frac{1-\alpha_{\rm mbl}}{\alpha_{\rm gr}-\alpha_{\rm mbl}}\right)^{\frac{\alpha_{\rm gr}-\alpha_{\rm mbl}}{1-\alpha_{\rm gr}}}\frac{1-\alpha_{\rm mbl}}{m_0} \left(\frac{m_0}{m_{\infty}}\right)^{1-\alpha_{\rm mbl}}\ .
\end{equation}
and the choice of the developmental velocity $K$ can be translated into a choice of a developmental velocity factor $0<\lambda_K\leq1$ such that 
\begin{equation}
    K=\lambda_K\, \gamma\delta\kappa_0\,
    Q
    \ , 
\end{equation}
which relates developmental velocity with metabolic scaling, growth scaling, and the characteristic mass scales of the system.

When $\alpha_{\rm gr}\leq \alpha_{\rm mbl}$ or the maximum is at $m(0),$ we have
\begin{equation}
K
\leq
\gamma\delta\kappa_0\;
\frac{(1-\alpha_{\rm gr})\,{m_0}^{-\alpha_{\rm mbl}}\,m(0)^{\alpha_{\rm mbl}-\alpha_{\rm gr}}\,{m_\infty}^{\alpha_{\rm gr}-1}}{
1-\left(\frac{m(0)}{m_\infty}\right)^{1-\alpha_{\rm gr}}
}\ ,
\end{equation}
yielding a constraint on the developmental velocity.
\subsubsection{Illustrative Examples}
To illustrate the concept with concrete examples, we compare two species  (see Figure~\ref{fig:pgr_fig}) with markedly different life histories: the dwarf goby (\textit{Eviota queenslandica}), a short-lived species exhibiting nearly linear growth, and the Greenland shark (\textit{Somniosus microcephalus}), the longest-lived vertebrate known, whose lifespan can extend to several hundred years~\cite{nielsen2016}. Bertalanffy coefficients have been measured for many fish species, and are available online~\cite{fishbase}. For dwarf goby, we use the parameters from~\cite{morais2018}: $L_{\infty}=2.98$ cm and $K=7.227$ year$^{-1}$ with maximum length of $2.6$ cm. The parameter $t_0$ is not reported, so we set it to $-0.0137$ years, such that the maximum length of the species is reached at an age of
99 days~\cite{depczynski2006}. The scaling relationship $m=aL^b$ was estimated~\cite{morais2018,froese2014} as $a=0.00891$ and $b=3.07,$ i.e.~in our notation $\beta=1/3.07\approx0.33.$ Furthermore, $\alpha=0$ and $\alpha_{\rm gr}$ is then $1-\beta\approx 0.67$ (see Eq.~\eqref{eq:grvsalpha_beta}).
For Greenland shark, the reported parameters are: $L_{\infty}=527.1$cm, $K=0.008$ year$^{-1}$ and $t_0=-10.7$ year. The mass-length relationship is described by $a=0.00135$ and $b=3.311$~\cite{fishbase}, i.e.~in our notation $\beta\approx0.3.$

To estimate the investment ratio $p_{\rm{gr}}$, we would need to specify the coefficients $\gamma, \delta$ and $\kappa_0$, which represent the metabolic-energy-to-mass conversion factor, the ratio of surplus to baseline metabolic rate, and the baseline metabolic rate at the reference mass $m_0$ (set to $0.1$ g), respectively. 
As a reliable estimation of the required parameters is non-trivial and beyond the scope of this study, we plot $p_{\rm gr} /\hat{p}_{\rm gr}$, where $\hat{p}_{\rm gr}$ is the maximal relative growth investment. This choice 
allows us to illustrate qualitative behaviour rather than to provide quantitative predictions.

The resulting time-dependent growth investment ratio $p_{\rm{gr}}$ is shown in Figure~\ref{fig:pgr_fig} and represents the fraction of available energy per unit time allocated to growth. As derived above (see Eq.~\eqref{eq:pgr_max}), when $\alpha_{\rm mbl} < \alpha_{\rm{gr}}$, the investment ratio is non-monotonic, whereas in all other cases it decreases monotonically over time, where species invest the highest portion of the surplus metabolic rate early in life. 

Keeping the phenomenological growth curve invariant, following
the von Bertalanffy fit, allows us to analyse how different metabolic scaling exponents shape the growth investment function $p_{\rm gr}$ through the relation $\alpha_{\rm met}=\alpha_{\rm gr}-\alpha_{\rm mbl}$.  If metabolic scaling is decoupled from growth scaling in this way, identical growth trajectories may correspond to different metabolic regimes, implying that growth curves alone do not uniquely determine the metabolic scaling exponent.

The qualitative regimes of $p_{\rm gr}$ are consistent across both species despite large differences in lifespan and body size, highlighting that allocation dynamics in our model are governed primarily by scaling relationships rather than absolute biological timescales. Notably, parameter values near $\alpha_{\rm mbl} \approx 0.9$, which are consistent with empirical intraspecific estimates for many fish species~\cite{jerde2019}, fall within the monotonic regime and produce rapidly declining growth investment over ontogeny. However, for the goby, it declines to a higher terminal value.
\section{Discussion}
A central result is the relation between the mass-based growth exponent $\alpha_{\rm gr}$, the baseline metabolic scaling exponent $\alpha_{\rm mbl}$, the length-based von Bertalanffy exponent $\alpha$, the geometric exponent $\beta$ and the metabolic investment exponent $\alpha_{\rm met}\equiv\alpha_{\rm gr}-\alpha_{\rm mbl}$ (see Table~\ref{tab:symbols}).
These exponents are not independent but connected by the equation
\begin{equation}
   \alpha_{\rm met}=1-\beta(1-\alpha)-\alpha_{\rm mbl} \ , 
\end{equation}
in a way reminiscent of exponent relations in the physics of critical phenomena.
For instance, the standard values $\alpha_{\rm mbl}=3/4$, {${\alpha=0}$}
and $\beta=1/3$ imply $\alpha_{\rm met}=-1/12$, i.e.~$\alpha_{\rm gr}=2/3$ in which case the investment ratio $p_{\rm gr}$ is a decreasing function, and the species spends a decreasing portion of metabolic energy on growth throughout its life. This corresponds to the conventional regime in which organisms prioritise growth early in ontogeny and progressively shift energy toward other functions.

If $\alpha_{\rm met}>0$ and if the maximiser $\hat m$ lies within the admissible mass range, then the relative metabolic investment into growth $p_{\rm gr}$ is not a purely monotonically decreasing function but increases until $m$ reaches the maximiser $\hat m$ to then monotonically decrease. Consequently, early in ontogeny, a larger fraction of surplus metabolic rate $(1-p_{\rm gr})\,\delta\varepsilon_0(m)$ may be allocated to non-growth processes beyond baseline maintenance. These channels of energy investments may vary widely across kingdoms and even species.  

Our model relies on several simplifying assumptions. First, we treat the energy-to-mass conversion factor $\gamma$ as constant throughout the lifetime, although in reality it may vary due to changes in tissue composition and the production
of different types of biomass during ontogeny. 
Second, we adopt a fixed geometric description and a fixed value of $\delta$, although organismal shape and the energetic expenditures captured by $\delta$ may change nontrivially during development. Furthermore, we assume that total metabolic energy and baseline metabolic requirements scale with the same exponent. Finally, 
baseline maintenance expenditure is assumed to follow a power-law scaling with exponent $\alpha_{\rm mbl}$ throughout the whole lifetime. Although biologically simplified, these assumptions allow us to isolate the relationships among the scaling exponents that constitute the focus of this work.
\section{Conclusions}
In this study, we linked von Bertalanffy growth dynamics to the metabolic energy allocation model. 
We decoupled growth scaling from metabolic scaling and showed how identical growth trajectories may arise from a variety of
energy allocation  
strategies.
We derived an explicit relationship linking the metabolic investment exponent to the exponents in the von Bertalanffy growth function and the geometric scaling exponent, and analysed the qualitative behaviour of the allocation function~$p_{\rm gr}$.
The biological requirement $0\leq p_{\rm gr}\leq1$ then constrains the compatible
combinations of growth scaling, metabolic scaling, geometric scaling, and
developmental velocity.

While prior models describe growth through balance among anabolic input, maintenance, reproduction, reserves, or metabolic constraints~\cite{kooijman2010,hou2008}, our framework emphasises growth as an allocation problem.
We assume total metabolic energy is partitioned among baseline metabolism, growth, reproduction, and other functions. Growth is driven by energy that scales allometrically with body size and by a species-specific program $p_{\rm gr}$, which determines what portion of surplus
metabolic rate (beyond baseline) is devoted to growth. 
Our model characterises how scaling relationships shape the allocation of energy to growth.
This approach offers a physical energy-based perspective grounded in metabolic scaling principles and the partitioning of physiological energy. The allocation function $p_{\rm gr}$ thus makes explicit the link between metabolic scaling, growth dynamics, and energy partitioning within a single framework.
\section*{Acknowledgments}
This project has received funding from the European Union’s Horizon 2020 research and innovation programme under grant agreement No.~955708. HK also acknowledges support from the Slovak Grant Agency for Science (VEGA 2/0094/26). 
\bibliography{references}

@article{bertalanffy1957,
  title={Quantitative laws in metabolism and growth},
  author={von Bertalanffy, Ludwig},
  journal={The Quarterly Review of Biology},
  volume={32},
  number={3},
  pages={217--231},
  year={1957},
  publisher={American Institute of Biological Sciences},
  doi={10.1086/401873}
}

@article{brody1946,
  title={Bioenergetics and growth},
  author={Brody, Samuel and Lardy, Henry A},
  journal={The Journal of Physical Chemistry},
  volume={50},
  number={2},
  pages={168--169},
  year={1946},
doi = {10.1021/j150446a008}
}

@article{depczynski2006,
 ISSN = {00129658, 19399170},
 URL = {http://www.jstor.org/stable/20069341},
 author = {Martial Depczynski and David R. Bellwood},
 journal = {Ecology},
 number = {12},
 pages = {3119--3127},
 publisher = {Ecological Society of America},
 title = {Extremes, Plasticity, and Invariance in Vertebrate Life History Traits: Insights from Coral Reef Fishes},
 urldate = {2025-11-18},
 volume = {87},
 year = {2006}
}

@article{dodds2001,
  title={Re-examination of the “3/4-law” of metabolism},
  author={Dodds, Peter Sheridan and Rothman, Daniel H and Weitz, Joshua S},
  journal={Journal of Theoretical Biology},
  volume={209},
  number={1},
  pages={9--27},
  year={2001},
  publisher={Elsevier},
  doi = {10.1006/jtbi.2000.2238}
}

@article{dmitriew2011,
author = {Dmitriew, Caitlin M.},
title = {The evolution of growth trajectories: what limits growth rate?},
journal = {Biological Reviews},
volume = {86},
number = {1},
pages = {97-116},
doi = {https://doi.org/10.1111/j.1469-185X.2010.00136.x},
url = {https://onlinelibrary.wiley.com/doi/abs/10.1111/j.1469-185X.2010.00136.x},
year = {2011}
}

@article{feldman1983,
  title={The 3/4 mass exponent for energy metabolism is not a statistical artifact},
  author={Feldman, Henry A and McMahon, Thomas A},
  journal={Respiration Physiology},
  volume={52},
  number={2},
  pages={149--163},
  year={1983},
  publisher={Elsevier},
  doi = {10.1016/0034-5687(83)90002-6}
}

@misc{fishbase,
author = {Froese, Rainer and Pauly, Daniel},
title = {Fish{B}ase},
howpublished = {World Wide Web electronic publication, www.fishbase.org  ( 04/2025 )},
url = {https://www.fishbase.org},
year = {2025}
}

@article{froese2014,
author = {Froese, R. and Thorson, J. T. and Reyes Jr, R. B.},
title = {A Bayesian approach for estimating length-weight relationships in fishes},
journal = {Journal of Applied Ichthyology},
volume = {30},
number = {1},
pages = {78-85},
doi = {https://doi.org/10.1111/jai.12299},
url = {https://onlinelibrary.wiley.com/doi/abs/10.1111/jai.12299},

year = {2014}
}

@article{glazier2005,
author = {Glazier, Douglas S.},
title = {Beyond the ‘3/4-power law’: variation in the intra-and interspecific scaling of metabolic rate in animals},
journal = {Biological Reviews},
volume = {80},
number = {4},
pages = {611-662},
doi = {https://doi.org/10.1017/S1464793105006834},
year = {2005}
}

@article{glazier2022,
    author = {Glazier, Douglas S.},
    title = {Variable metabolic scaling breaks the law: from ‘Newtonian’ to ‘Darwinian’ approaches},
    journal = {Proceedings of the Royal Society B: Biological Sciences},
    volume = {289},
    number = {1985},
    pages = {20221605},
    year = {2022},
    month = {10},
    issn = {0962-8452},
    doi = {10.1098/rspb.2022.1605},
    url = {https://doi.org/10.1098/rspb.2022.1605},
}

@article{halbersleben1922,
title = {Relation of Egg Weight to Chick Weight at Hatching.},
journal = {Poultry Science},
volume = {1},
number = {4},
pages = {143-144},
year = {1922},
issn = {0032-5791},
doi = {https://doi.org/10.3382/ps.0010143},
url = {https://www.sciencedirect.com/science/article/pii/S0032579119567311},
author = {D.L. Halbersleben and F.E. Mussehl}
}

@article{heusner1982,
title = {Energy metabolism and body size I. Is the 0.75 mass exponent of Kleiber's equation a statistical artifact?},
journal = {Respiration Physiology},
volume = {48},
number = {1},
pages = {1-12},
year = {1982},
issn = {0034-5687},
doi = {https://doi.org/10.1016/0034-5687(82)90046-9},
url = {https://www.sciencedirect.com/science/article/pii/0034568782900469},
author = {A.A. Heusner}
}

@article{hou2008,
  title={Energy uptake and allocation during ontogeny},
  author={Hou, Chen and Zuo, Wenyun and Moses, Melanie E and Woodruff, William H and Brown, James H and West, Geoffrey B},
  journal={Science},
  volume={322},
  number={5902},
  pages={736--739},
  year={2008},
  publisher={American Association for the Advancement of Science},
  doi = {10.1126/science.1162302}
}

@article{jerde2019,
  title={Strong evidence for an intraspecific metabolic scaling coefficient near 0.89 in fish},
  author={Jerde, Christopher L. and Kraskura, Krista and Eliason, Erika J. and Csik, Samantha R. and Stier, Adrian C. and Taper, Mark L.},
  journal={Frontiers in Physiology},
  volume={10},
  pages={1166},
  year={2019},
  publisher={Frontiers Media SA},
  doi = {10.3389/fphys.2019.01166}
}

@article{kearney2021,
  title={What is the status of metabolic theory one century after {P}\"{u}tter invented the von {B}ertalanffy growth curve?},
  author={Kearney, Michael R},
  journal={Biological Reviews},
  volume={96},
  number={2},
  pages={557--575},
  year={2021},
  publisher={Wiley Online Library},
  doi={doi: 10.1111/brv.12668}
}

@article{kleiber1932,
  title={Body size and metabolism},
  author={Kleiber, Max},
  journal={Hilgardia},
  volume={6},
  number={11},
  pages={315--353},
  year={1932},
  publisher={University of California, Agriculture and Natural Resources},
  doi = {10.3733/hilg.v06n11p315}
}

@book{kooijman2010,
  title={Dynamic energy budget theory for metabolic organisation},
  author={Kooijman, Sebastiaan Adriaan Louis Maria},
  year={2010},
  publisher={Cambridge University Press},
  doi = {10.1017/CBO9780511805400}
}

@article{lee2020,
  title={A new framework for growth curve fitting based on the von {Bertalanffy} Growth Function},
  author={Lee, Laura and Atkinson, David and Hirst, Andrew G and Cornell, Stephen J},
  journal={Scientific Reports},
  volume={10},
  number={1},
  pages={7953},
  year={2020},
  publisher={Nature Publishing Group UK London},
  doi = {10.1038/s41598-020-64839-y}
}

@article{makarieva2004,
  title={Ontogenetic growth: models and theory},
  author={Makarieva, Anastassia M and Gorshkov, Victor G and Li, Bai-Lian},
  journal={Ecological Modelling},
  volume={176},
  number={1-2},
  pages={15--26},
  year={2004},
  publisher={Elsevier},
  doi = {10.1016/j.ecolmodel.2003.09.037}
}

@article{morais2018,
author = {Morais, Renato A. and Bellwood, David R.},
title = {Global drivers of reef fish growth},
journal = {Fish and Fisheries},
volume = {19},
number = {5},
pages = {874-889},
doi = {https://doi.org/10.1111/faf.12297},
url = {https://onlinelibrary.wiley.com/doi/abs/10.1111/faf.12297},
year = {2018}
}

@article{nielsen2016,
  title={Eye lens radiocarbon reveals centuries of longevity in the {Greenland shark (Somniosus microcephalus)}},
  author={Nielsen, Julius and Hedeholm, Rasmus B and Heinemeier, Jan and Bushnell, Peter G and Christiansen, J{\o}rgen S and Olsen, Jesper and Ramsey, Christopher Bronk and Brill, Richard W and Simon, Malene and Steffensen, Kirstine F. and Steffensen, John F. },
  journal={Science},
  volume={353},
  number={6300},
  pages={702--704},
  year={2016},
  publisher={American Association for the Advancement of Science},
 doi = {10.1126/science.aaf1703}
}

@book{peters1986, place={Cambridge}, series={Cambridge Studies in Ecology}, title={The Ecological Implications of Body Size}, publisher={Cambridge University Press}, author={Peters, Robert Henry}, year={1983}, collection={Cambridge Studies in Ecology}}

@article{putter1920,
  title={{Studien {\"u}ber physiologische {\"A}hnlichkeit VI. Wachstums{\"a}hnlichkeiten}},
  author={P{\"u}tter, August},
  journal={Pfl{\"u}ger's Archiv f{\"u}r die Gesamte Physiologie des Menschen und der Tiere},
  volume={180},
  number={1},
  pages={298--340},
  year={1920},
  publisher={Springer-Verlag Berlin/Heidelberg},
  doi={https://doi.org/10.1007/BF01755094}
}

@article{renner2018,
  title={On the exponent in the von {B}ertalanffy growth model},
  author={Renner-Martin, Katharina and Brunner, Norbert and K{\"u}hleitner, Manfred and Nowak, Werner Georg and Scheicher, Klaus},
  journal={PeerJ},
  volume={6},
  pages={e4205},
  year={2018},
  publisher={PeerJ Inc.},
  doi = {10.7717/peerj.4205}
}

@article{rubner1883,
  title={{\"U}ber den einfluss der k{\"o}rpergr{\"o}sse auf stoff-und kraftwechsel},
  author={Rubner, Max},
  journal={Zeitschrift f{\"u}r Biologie},
  volume={19},
  pages={536},
  year={1883}
}

@article{sarrus1838,
  title={Rapport sur une m{\'e}moire adress{\'e} {\'a} l’Acad{\'e}mic royale de {M\'e}decine},
  author={Sarrus, F and Rameaux, J},
  journal={Bulletin de {l'Académie Royale de Médecine de Paris}},
  volume={3},
  pages={1094--1100},
  year={1838}
}

@article{savage2004,
  title={The predominance of quarter-power scaling in biology},
  author={Savage, Van M and Gillooly, James F and Woodruff, William H and West, Geoffrey B and Allen, Andrew P and Enquist, Brian J and Brown, James H},
  journal={Functional Ecology},
  volume={18},
  number={2},
  pages={257--282},
  year={2004},
  publisher={Wiley Online Library},
  doi = {10.1111/j.0269-8463.2004.00856.x}
}

@article{sibly2020,
  title={Toward a physiological explanation of juvenile growth curves},
  author={Sibly, RM and Brown, JH},
  journal={Journal of Zoology},
  volume={311},
  number={4},
  pages={286--290},
  year={2020},
  publisher={Wiley Online Library},
doi = {10.1111/jzo.12770}
}

@misc{stiefel2022,
  title        = {What Determines the Minimum Body Size for Vertebrates?},
  author       = {Stiefel, Klaus and Bucol, Abner A.},
  year         = {2022},
  doi = {doi.org/10.32942/osf.io/k72at},
  publisher = {Preprint available at EcoEvoRxiv},
}

@article{tjorve2010,
  title={Shapes and functions of bird-growth models: how to characterise chick postnatal growth},
  author={Tj{\o}rve, Kathleen M. C. and Tj{\o}rve, Even},
  journal={Zoology},
  volume={113},
  number={6},
  pages={326--333},
  year={2010},
  publisher={Elsevier},
  doi = {10.1016/j.zool.2010.05.003}
}

@article{west1997,
  title={A general model for the origin of allometric scaling laws in biology},
  author={West, Geoffrey B and Brown, James H and Enquist, Brian J},
  journal={Science},
  volume={276},
  number={5309},
  pages={122--126},
  year={1997},
  publisher={American Association for the Advancement of Science},
  doi = {10.1126/science.276.5309.122}
}

@article{west2001,
  title={A general model for ontogenetic growth},
  author={West, Geoffrey B and Brown, James H and Enquist, Brian J},
  journal={Nature},
  volume={413},
  number={6856},
  pages={628--631},
  year={2001},
  publisher={Nature Publishing Group UK London},
  doi = {10.1038/35098076}
}

@article{white2003,
  title={Mammalian basal metabolic rate is proportional to body mass 2/3},
  author={White, Craig R and Seymour, Roger S},
  journal={Proceedings of the National Academy of Sciences},
  volume={100},
  number={7},
  pages={4046--4049},
  year={2003},
  publisher={National Acad Sciences},
  doi = {10.1073/pnas.0436428100}
}
\end{document}